\documentclass[sigconf]{acmart}

\AtBeginDocument{%
  }

\copyrightyear{}
\acmYear{}
\acmDOI{}

\acmConference[]{}{}{}
\acmISBN{}

\usepackage{bmpsize}

\usepackage{times}    % comment if you want LaTeX's default font
\usepackage{url}      % llt: nicely formatted URLs
\usepackage{amsmath}
\usepackage{bm}
\usepackage{stmaryrd}
\usepackage{here}
\usepackage{url}

\usepackage{subcaption}
\usepackage{multirow}
\usepackage{booktabs}
\usepackage{bm}
\usepackage{ascmac}

\renewcommand\footnotetextcopyrightpermission[1]{} % コピーライト情報を非表示

\begin{document}

%%
%% The "title" command has an optional parameter,
%% allowing the author to define a "short title" to be used in page headers.
\title{DisasterNeedFinder: Understanding the Information Needs in the 2024 Noto Earthquake (Comprehensive Explanation)}

%%
%% The "author" command and its associated commands are used to define
%% the authors and their affiliations.
%% Of note is the shared affiliation of the first two authors, and the
%% "authornote" and "authornotemark" commands
%% used to denote shared contribution to the research.
\author{Kota Tsubouchi}
\affiliation{%
  \institution{LY Corporation}
  \city{Tokyo}
  \country{Japan}
}
\email{ktsubouc@lycorp.co.jp}

\author{Shuji Yamaguchi}
\affiliation{%
  \institution{LY Corporation}
  \city{Tokyo}
  \country{Japan}
}
\email{shyamagu@lycorp.co.jp}

\author{Keijirou Saitou}
\affiliation{%
  \institution{Japan Broadcasting Corporation}
  \city{Tokyo}
  \country{Japan}
}
\email{saitou.k-oa@nhk.or.jp}

\author{Akihisa Soemori}
\affiliation{%
  \institution{NHK Global Media Servises}
  \city{Tokyo}
  \country{Japan}
}
\email{soemori-a@nhk-g.co.jp}

\author{Masato Morita}
\affiliation{%
  \institution{NHK Global Media Servises}
  \city{Tokyo}
  \country{Japan}
}
\email{morita-ma@nhk-g.co.jp}

\author{Shigeki Asou}
\affiliation{%
  \institution{Japan Broadcasting Corporation}
  \city{Tokyo}
  \country{Japan}
}
\email{asou.s-cm@nhk.or.jp}

%%
%% By default, the full list of authors will be used in the page
%% headers. Often, this list is too long, and will overlap
%% other information printed in the page headers. This command allows
%% the author to define a more concise list
%% of authors' names for this purpose.
\renewcommand{\shortauthors}{Tsubouchi et al.}

%%
%% The abstract is a short summary of the work to be presented in the
%% article.
\begin{abstract}

We propose and demonstrate the DisasterNeedFinder framework in order to provide appropriate information support for the Noto Peninsula Earthquake. In the event of a large-scale disaster, it is essential to accurately capture the ever-changing information needs. However, it is difficult to obtain appropriate information from the chaotic situation on the ground. Therefore, as a data-driven approach, we aim to pick up precise information needs at the site by integrally analyzing the location information of disaster victims and search information. It is difficult to make a clear estimation of information needs by just analyzing search history information in disaster areas, due to the large amount of noise and the small number of users.
Therefore, the idea of assuming that the magnitude of information needs is not the volume of searches, but the degree of abnormalities in searches, enables an appropriate understanding of the information needs of the disaster victims.
DNF has been continuously clarifying the information needs of disaster areas since the disaster strike, and has been recognized as a new approach to support disaster areas by being featured in the major Japanese media on several occasions.
\end{abstract}

%%
%% The code below is generated by the tool at http://dl.acm.org/ccs.cfm.
%% Please copy and paste the code instead of the example below.
%%


%%
%% Keywords. The author(s) should pick words that accurately describe
%% the work being presented. Separate the keywords with commas.
%% A "teaser" image appears between the author and affiliation
%% information and the body of the document, and typically spans the
%% page.

%\received{7 June 2024}

%%
%% This command processes the author and affiliation and title
%% information and builds the first part of the formatted document.
\maketitle

\pagestyle{plain}

\begin{figure}[h]
 \includegraphics[width=1.0\linewidth]{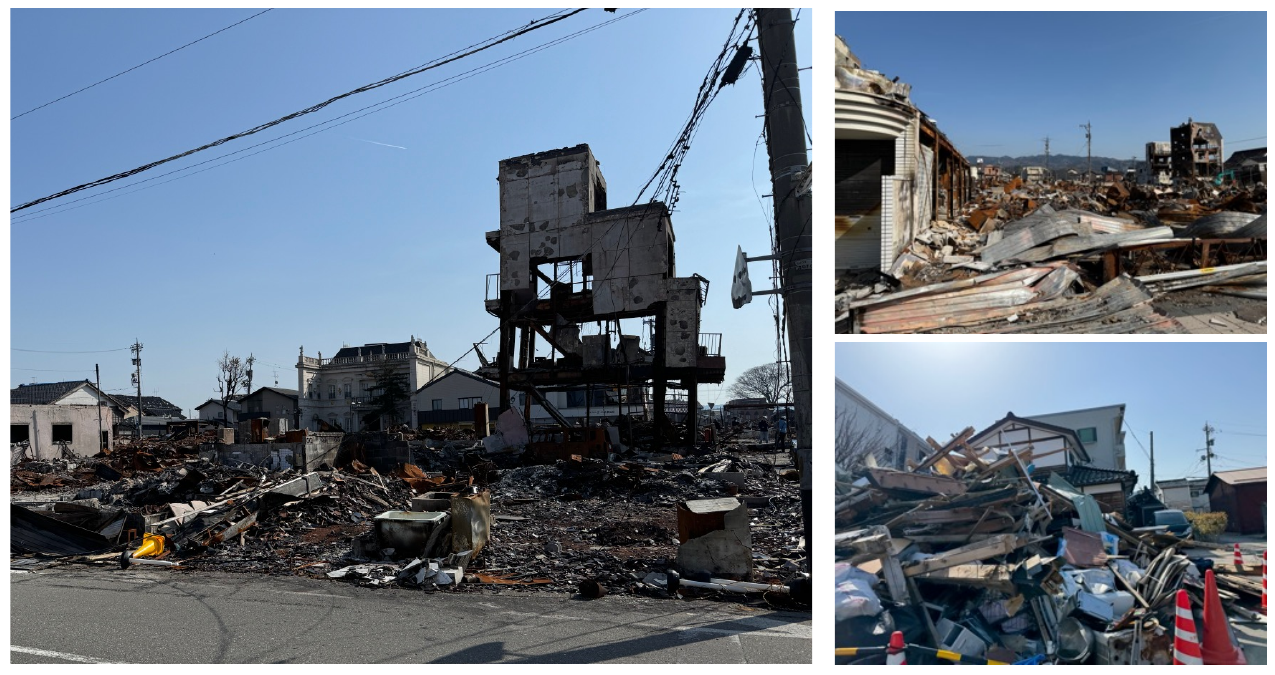}
 \caption{A scenery of the disaster area in April 2024 (photo by Kazuto Ataka)}
 \label{fig:photo_noto}
\end{figure}

\section{Introduction}
{\bf 2024 Noto Peninsula Earthquake (hereafter Noto Earthquake)} - {\it An inland crustal earthquake occurred 16 km beneath the Noto Peninsula in Ishikawa Prefecture, Japan, on January 1, 2024. According to the Japan Meteorological Agency, the magnitude of this earthquake was 7.6, making it one of the few inland earthquakes of its magnitude in Japan. The main quake triggered tsunamis in a wide area along the coast of the Sea of Japan, including outside Japan, as well as landslides, fires, and liquefaction in many areas. The earthquake caused a series of collapsed houses, resulting in more than 200 fatalities and extensive damage throughout the Hokuriku region, especially in the Okunoto area. Because the earthquake occurred on a peninsula, the transportation network was cut off, and rescue operations by the Self-Defense Forces were difficult. In addition, lifelines such as electricity, water, and gas were damaged, and in many areas they have not been restored even after more than six months. (Figure \ref{fig:photo_noto} depicts the scenery of Wajima City four months after the disaster.)}

\vspace{10px}
When a large-scale disaster such as the Noto earthquake strikes, the information needs of the community are diverse and change from moment to moment. Especially in the immediate days following a disaster, the national government, local governments, and the Self-Defense Forces do not have sufficient support systems in place, and the resources available for restoration assistance are limited. Under such chaotic situations, it is needless to say that it is essential to grasp the dynamics of the disaster area, to accurately grasp the ever-changing needs on the ground, and to take appropriate support actions in response to those needs. Research and practical use of location information obtained from smartphones to understand the movements of people in disaster-stricken areas has been conducted extensively. On the other hand, the understanding of information needs in disaster-stricken areas is still insufficient.

Studies have often looked at analyzing social networking services such as Twitter \cite{Liu:2016} and trend analysis \cite{carneiro2009google} using search queries and purchase histories as a way to understand the information needs that change from moment to moment during a large-scale disaster or during the recovery process. 
However, there are three challenges to these studies.
These are unstable access conditions, the impact of media exposure, and weak signals in areas with a small number of users. First, the unstable access situation means that the time and timing of user use in the affected areas will be very different from what it has been in the past. For example, users may only use the minimum amount of communication, as they place the highest priority on securing power, they may not have time to enjoy entertainment, or they may not have time to operate their smartphones to secure a minimum level of food, clothing, and shelter. An analysis that takes these instabilities into account is needed.
Secondly, the larger the scale of the disaster, the more news reports are likely to be broadcast. In general, when a disaster is reported in the news, people nationwide become more aware of it, and a certain number of people start to search for it or tweet about it to give notice of their perspective \cite{cassa2013twitter}. These increased information needs are merely a response to the news reports, and cannot be considered an increase in the information needs of the disaster area. Factors such as news reports should be eliminated in order to detect the information needs required in the disaster areas. 
Finally, if disaster areas are areas with a small number of users, the signals for detecting information needs are often small and difficult to estimate. To begin with, in cases such as the Noto earthquake, the population of disaster areas is small compared to the rest of the country, thus making it difficult for many SNS and web search services to separate information, such as whether it is from people in the disaster areas or people outside the disaster areas. In such a situation, it is difficult to pick up and organize only signals from disaster areas.

To address these issues, DisasterNeedFinder (DNF) framework that aims to extract information needs in disaster areas through integrated analysis of search history information and location information is proposed. The proposed DNF framework is characterized by both data collection and analysis processes. First, in the data collection process, not only search queries are collected, but also users' location information is collected. This makes it possible to determine whether the search query originated from the target area or not. The analysis process also uses a learning method with automatic generation of stopwords to offset the influence of news coverage and unique stopwords to describe the target region. Furthermore, the idea of defining the intensity of information needs as the anomaly in the number of queries retrieved, rather than the total number of queries, has made it possible to accurately extract the ever-changing information needs of disaster areas, even when the number of users is small and access is unstable.

The DNF framework was used during the Noto earthquake and demonstrated its effectiveness through the actual disaster, including coverage in Japan's major media and accurate identification of local information needs. This framework is an effective method not only for the Noto Peninsula earthquake, but also for many other large-scale disasters.

The contributions of this research are twofold:
\begin{itemize}
    \item The DNF framework was proposed to accurately grasp the ever-changing information needs of the field in the event of a large-scale disaster.
    \item The DNF framework was demonstrated and put to practical use in the Noto Peninsula earthquake, and the effectiveness of the framework was verified.
\end{itemize}

\section{Related Works}
\subsection{Information Management in Disasters}
Increasingly, location-based information management is being used in disaster management. Sun et al. \cite{sun:2020} review the use of AI in disaster management, covering its four phases — mitigation, preparedness, response, and recovery — across a total of 26 AI methods in 17 application areas.

Simulation of evacuation behavior, estimation of informal shelters, flood analysis, and disaster-specific recovery estimation are being conducted\cite{Chondrogiannis:2021,Herschelman:2019,Elnaz:2021,Xu:2023,Abdullahi:2018}.
Ochiai et al. \cite{Ochiai:2022} propose a method using autoencoder-based anomaly detection to identify non-designated evacuation shelters from cellular data, demonstrating improved accuracy over traditional methods.
Burke et al. \cite{Burke:2022} propose an interdisciplinary geospatial approach combining geosimulations, AI algorithms, and hydrodynamic modeling to assess flood risks under various urban scenarios, aiming to significantly impact urban growth research and community resilience.
Thakur et al. \cite{Thakur:2021} developed a scalable workflow to efficiently collect and harmonize data for effective disaster relief and recovery, emphasizing the importance of timely and accurate assessments and coordinated efforts across agencies within the first 24 hours post-disaster.

All of these studies observe the real behavior of people, and none of them analyze the inner life of disaster victims. This is the first study to estimate the inner information needs of disaster victims in terms of search queries.

\subsection{Location Data Utilization}
Many researchers have studied how user location information can be utilized. The position information obtained from the user can be utilized for various purposes. For instance, it can be used for making histories of location information as a life-log service \cite{Mehrotra:2017,Singh:2016}, modeling the behavior of individual users and urban dynamics \cite{Yabe:2017,Lee:2016,Fan:2016}, ride sharing \cite{Kim:2017,Yabe:2016,Asghari:2017}, and elaborating map and route information \cite{Wu:2016,Mridha:2017,Wang:2016}.

In particular, position information collected by smartphones have been used to analyze check-in histories in Foursquare \cite{Lichman:2016}, and position information such as geo tags attached to tweets have been used to analyze the behaviors of Twitter users \cite{Liu:2016}. However, in these cases, location information is not collected unless the user is checking in or posting a tweet. Such data is too sparse to be used to track a user's lifestyle.

Thanks to advances aimed at reducing their power consumption and improving battery efficiency, smartphones can be continually used to obtain location information. Zhu et al. \cite{zhu2015efficient} propose a new efficient and privacy preserving LBS query scheme in outsourced cloud. Wu et al. \cite{Wu:2017} constructed a calibration model that minimizes location estimation errors arising from variations in the strength of the radio field emitted from the base station. They used location information obtained frequently from users as ground truth. 
In other studies, high-frequency data were obtained from actual taxi probe data; the datasets had a high enough resolution for reproducing the trajectory of the user. Other such dense datasets have been acquired for bus \cite{Mazimpaka:2016} and car navigation applications \cite{Andersen:2017}. These datasets were obtained from actual services. As frequently acquiring position information leads to the battery consumption problem, the power in these cases was supplied by an in-vehicle battery. 

Some studies acquired location information based on position information continually obtained from the user's smartphone on basis of frequent trajectories. The study of Kim et al. \cite{Kim:2017} logged and tracked hiker's behavior when they were climbing hills, and they found it necessary to supplement the power with an auxiliary battery. On the other hand, Mehrotra et al. \cite{Mehrotra:2017} paid a reward to examinees and acquired position information with limited frequency. 

As listed above, there has been a lot of research on the analysis of location information. Most of the research is focused on understanding the context from the analysis of location information alone, but research to understand deeper context by combining it with some kind of historical information, as in this study, is gradually becoming popular.

\subsection{Combining Location and Other Data for Event Profiling}

Research that utilizes various data obtained from smartphones to profile events is actively conducted. Sakaki et al.\cite{sakaki2010earthquake} analyze user behavior during earthquake occurrences through the analysis of tweet data. While the case study in this research dates back to 2009, a majority of the tweet data at that time consisted of tweets with appended metadata containing GPS-derived location information. However, gradually from around 2013, this metadata indicating location information decreased, and currently, less than 1\% of all tweets are endowed with location information. Furthermore, within that 1\%, the majority of tweets comprise promotional tweets from companies, and location information is not attached as metadata to tweets from general users.

Since the metadata containing location information ceased to be appended to tweet data, studies that combine search queries and location information have gained attention. Jiang et al.\cite{jiang2022will} speculated whether users who searched for a specific facility would actually visit that facility, demonstrating a correlation between search behavior and actual offline behavior. Hisada et al.\cite{hisada2020surveillance} and Yabe et al.\cite{yabe2020non}, on the other hand, estimate areas with a high risk of COVID-19 cluster outbreaks by comprehensively analyzing search queries and location information. In these studies, they identify the risk of COVID-19 through search queries and track the movements of high-risk users to estimate areas with a high risk of cluster outbreaks.

Since the tweet data with location information attached is currently unavailable, the combination of search queries and location information is useful as data that integrates offline behavior and information needs of individuals. However, these studies focus solely on the movements and search behavior of individual users, without considering extracting information needs from a user group who are present in a specific area and participating in an event. This research, on the other hand, aims to extract information needs from a group of users who are present in a specific area and participating in an event, rather than focusing on individual users.

\section{DisasterNeedFinder Framework}

\begin{figure*}[t]
 \includegraphics[width=1.0\linewidth]{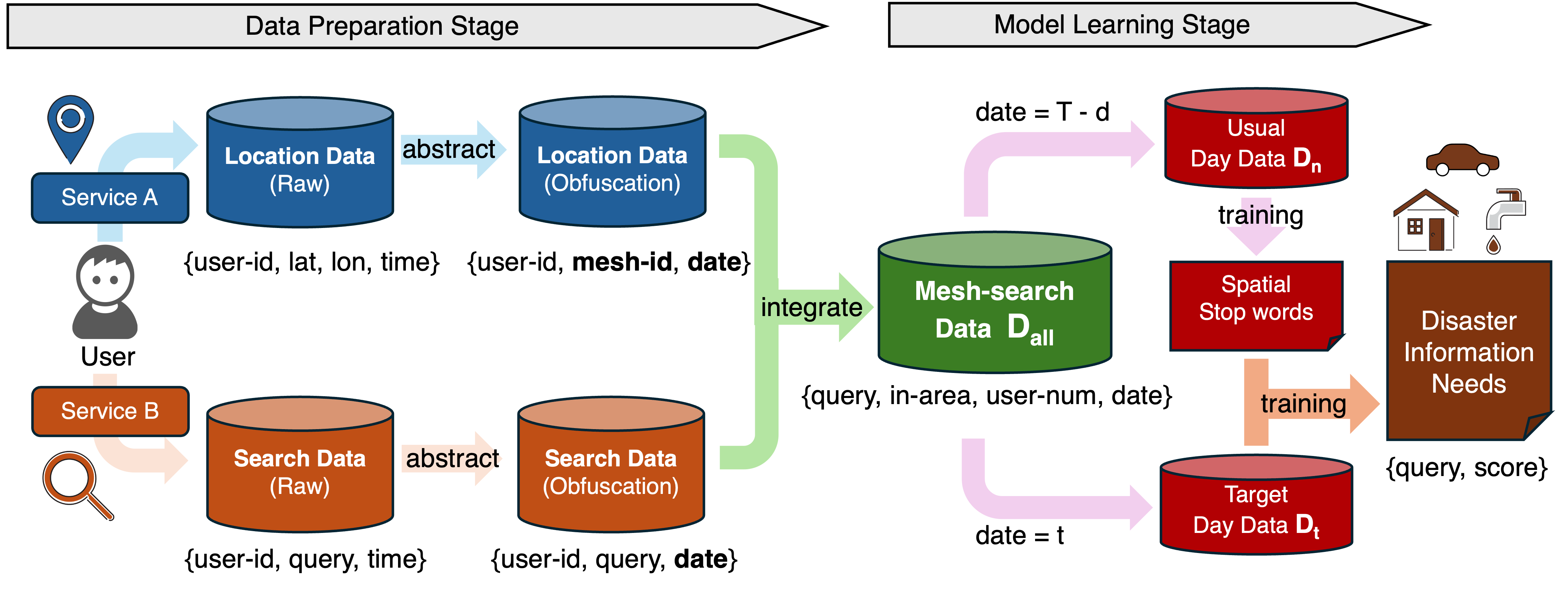}
 \caption{Overview of DisasterNeedFinder Framework}
 \label{fig:overview}
\end{figure*}

Figure \ref{fig:overview} shows an overview of the DNF framework.
The DNF framework consists of two stages: the Data Preparation Stage and the Model Learning Stage.
In the Data Preparation Stage, data is created for input into the Model Learning Stage.
Note that the important point is that the information needs of disaster victims can be obtained by integrating location information and search query information, but this is not the only way to achieve this goal.

\subsection{Overview of DNF Framework}
The DNF framework is required to reveal the information needs of disaster areas while protecting the privacy of users. The Model Learning Stage of the DNF requires information consisting of four elements: when, inside and outside the area, which information needs are being observed, and at what intensity.

The first step is to create the data set in the Data Preparation Stage.
LY Corporation is the largest Internet service provider in Japan, providing users with a number of Internet services such as Web search services and disaster prevention applications. LY Corporation stores search logs from web search services and user location logs from disaster prevention applications, maps, and other applications.
These raw logs contain a User-ID that identifies the user, and both logs can be analyzed in an integrated manner. At the same time, user privacy needs to be taken into consideration when creating data for analysis.
In the Model Learning Stage, we aim to extract information needs in disaster areas by focusing on three difficulties: 1) unstable access conditions, 2) exclusion of signals of nationwide interest such as news reports, and 3) weak signals specific to areas with a small number of users.

\subsection{Data Preparation Stage}

In the DNF framework, the Model Learning Stage requires information consisting of four elements: when, in which area (inside or outside the Noto Peninsula), the target search query, and by how many people. The Data Preparation Stage uses the user's location history and search history information to create this information while carefully considering the user's privacy.

LY Corporation is a company that provides various Internet services in Japan. The company offers a wide range of services, including web search, news applications, map applications, and disaster prevention applications. Users log in and use each service or application with a common ID. Therefore, logs from each service can be connected and analyzed.
Agreement to users is obtained not only from the consent provided by the OS, but also from consent to the privacy policy. In addition, location information is stored and managed through careful communication with users, for example, by obtaining individual consent for the use of location information separately from other data use.
We obtain query information from the usage history of web search services and user location information from applications that use location information such as disaster prevention and maps. This research uses these data for integrated analysis. In this process, obfuscation and k-anonymization are performed in advance to ensure the privacy of the user.

First, obfuscation process is explained below. Detailed location information is not required, and only information on whether the user is inside or outside the disaster area of the Noto Peninsula is used. First, we roughly estimate the user's location at the moment the user performs the search action and disambiguate the information into flags for whether the user is inside or outside the area. While general location information refers to latitude and longitude information, in this research, it is sufficient to determine whether the user is inside or outside of an area.
Moreover, since this research analyzes data on a daily basis, time information is also not necessary. Therefore, both the search query information and location information are dropped the time information to obfuscate the information into a daily unit.

Furthermore, k-anonymization is performed on the aggregated data. For example, if there is only one user who made a search query for something on a certain day, it could be information that identifies an individual. Therefore, the idea is to use only data with a certain number of observations in the same context (inside and outside the area, multiplied by the date and query) for the analysis. In this study, k is set to 9. The k-anonymized dataset is used as input information for the next stage.

These processes of obfuscation and k-anonymization aim to avoid using more information than necessary for the analysis process, which is very important in terms of user privacy.

\subsection{Model Learning Stage}
In the Model Learning Stage, training data created in the Data Preparation Stage is used for learning. In general, it is intuitive to use the total number of searches as the strength of information needs. However, even if queries such as those that are frequently searched on a daily basis appear at the top of the list, they are not the information needs required in disaster areas. Therefore, DNF considers the strength of information needs not as the number of searches, but as the increase or decrease in the number of searches compared to the usual number of searches. In other words, if query A, which is usually retrieved only 3 times, is retrieved 30 times due to the earthquake, it means that the information needs have been strengthened by 10 times.

To calculate the degree of abnormalities, a classification problem is solved using the search query as a feature, with data within the Noto Peninsula as a positive example and data outside the Noto Peninsula as a negative example. The positive and negative examples are aligned by undersampling, and learning is performed by linear regression. The reason for employing linear regression learning is that the weights assigned to the features are explicitly derived. These weights can be used to identify the dominant search queries to classify whether a user is in or out of the Noto Peninsula.
This method allows us to ignore the influence of queries that are suddenly retrieved on a national scale due to media coverage. This is because the contents of Breaking New are searched more frequently than usual, both inside and outside the Noto Peninsula.

On the other hand, this method often assigns a large weight to search queries that are unique to the Noto Peninsula, so it is necessary to take measures to deal with these queries. That is, for example, search queries such as “Noto Peninsula” or the name of a town in the Noto Peninsula are likely to be searched by people in the Noto Peninsula, and search queries are likely to be searched by people in the area. However, the information needs we wish to seek in this study are the ever-changing information needs of the people affected by the disaster in the Noto Peninsula and the people living there.
Therefore, this study introduces the “spatial stopwords” function into the learning process. In other words, the same kind of machine learning is performed at normal times before the earthquake, and the words that appear in the normal model at that time are treated as stopwords in the model at the time of the earthquake. For example, if the word “Noto Peninsula” is a characteristic search query that is usually searched for, it is used as a stopword, and the same word is devised so that it does not appear when learning the target day data.

By examining the weights assigned to each feature in the model trained in this way, we identified those with the highest values as strong local information needs.

\subsection{User Interface of DNF}

\begin{figure}[t]
 \includegraphics[width=\linewidth]{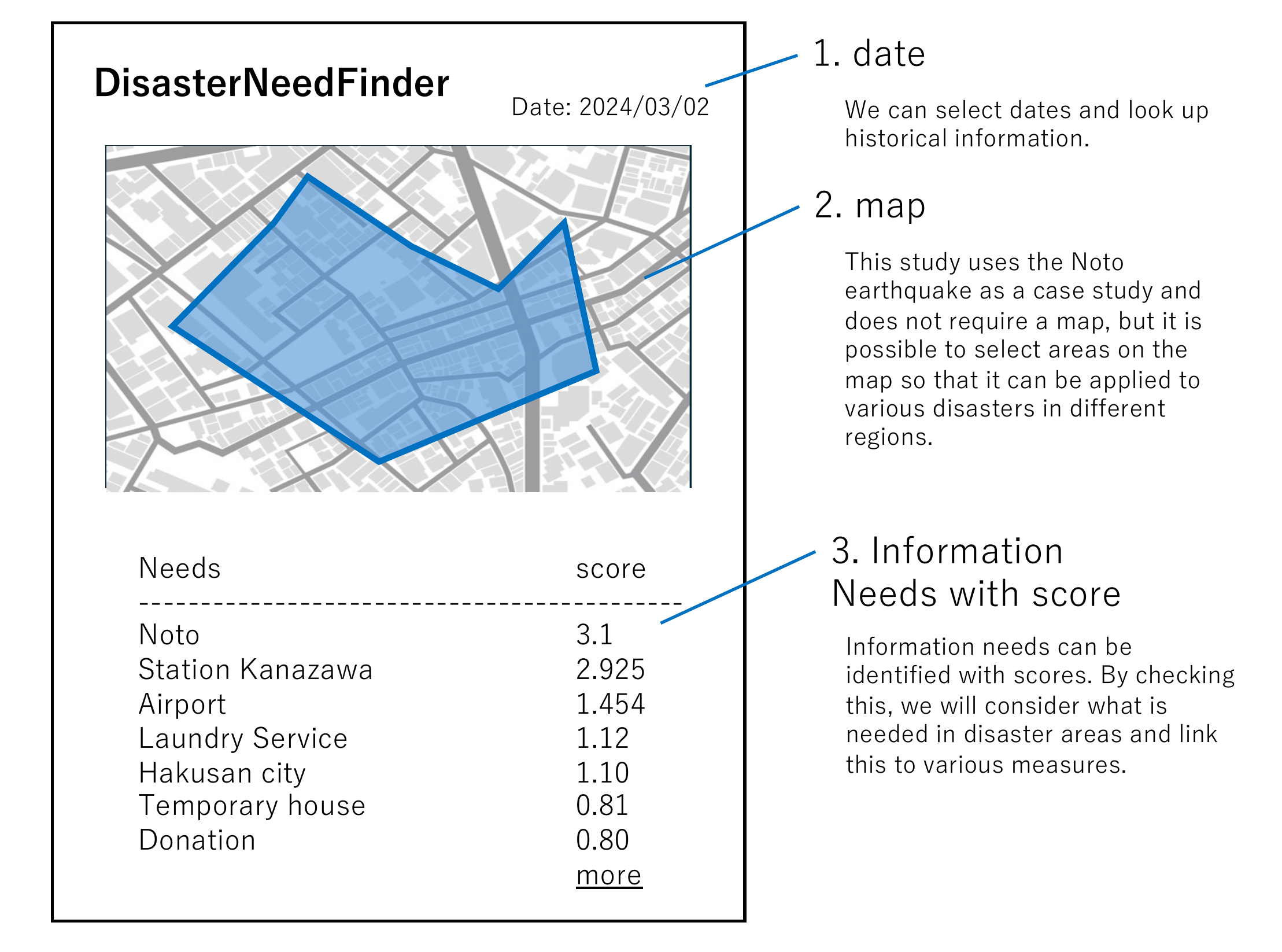}
 \caption{DisasterNeedFinder User Interface}
 \label{fig:if}
\end{figure}

A web application implementing the DNF framework was used to track the ever-changing information needs of the Noto earthquake disaster areas with a time lag of one day.
The reason for the one-day delay is that the unit of analysis is one-day aggregation.

The Noto Peninsula area was registered in advance as a positive example area and other areas as negative examples. When the date changes, data is collected, and after two stages, the information needs of the learned disaster areas can be viewed on the web application. This information enables users to see what they are currently looking for in disaster areas.
For confidentiality reasons, we are unable to show the actual application screen, so an illustration of the application screen is shown in Fig. \ref{fig:if}. As shown in the figure, this application is not specific to the Noto earthquake, but is designed to track the information needs of disaster areas in various disasters that have occurred in various regions.

\section{Validation in 2024 Noto Earthquake}
This system has been in operation since January 1, 2024, the day the Noto earthquake struck, and has continued to provide information.
The information needs of disaster areas provided by DNF were often highlighted in TV news reports and web articles, and information was provided throughout Japan.
As well as using the system as an actual service, we verified the effectiveness of this system and the validity of the estimated information needs. This chapter describes the results and the methods and results of the validation.

\subsection{Overview of the validation}

\begin{figure*}[p]
\rotatebox{90}{
\begin{minipage}{\textheight}\centering
\includegraphics[width=0.97\linewidth]{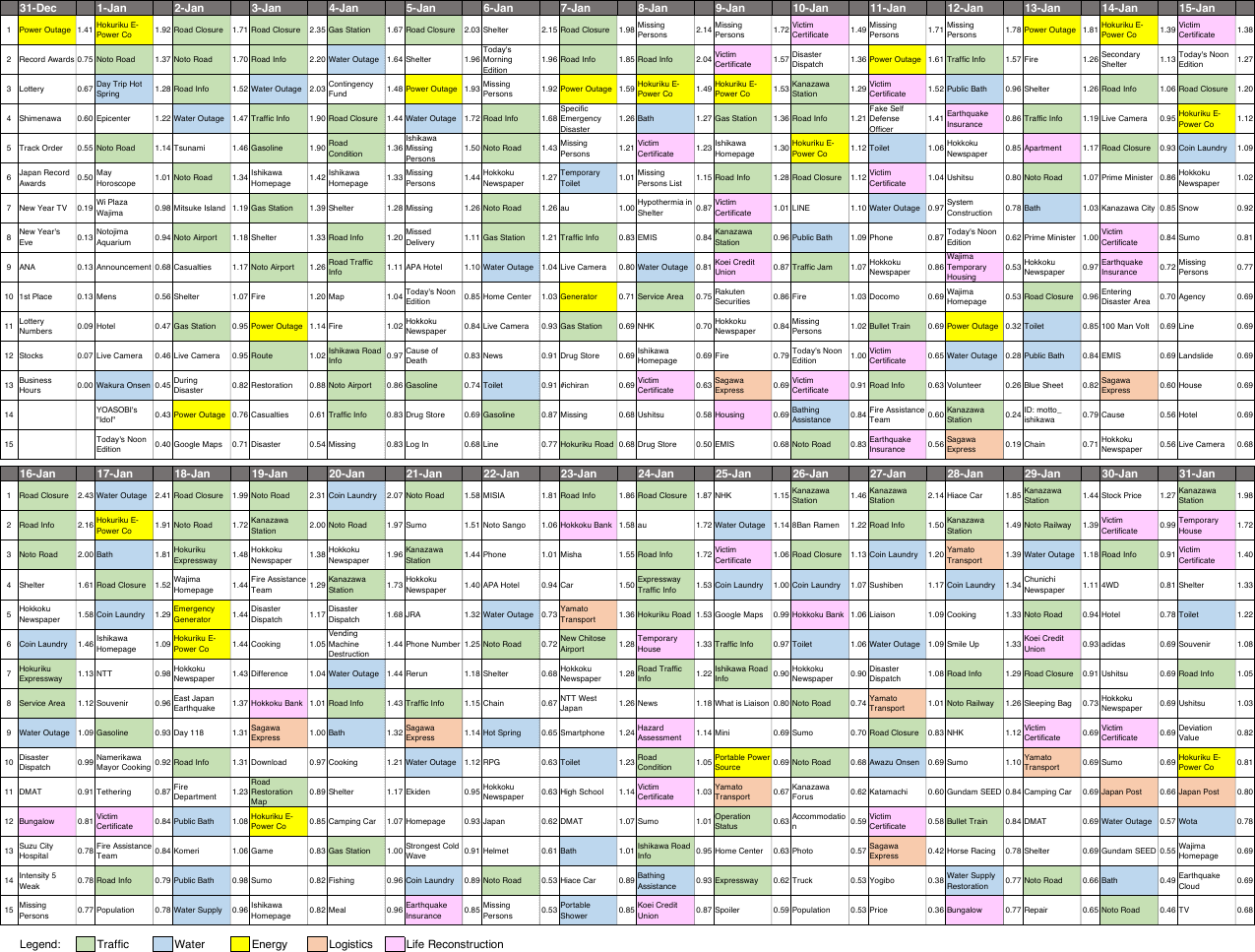}
\caption{The visualization of disaster information needs to be changed for the period of one month after Noto earthquake in Noto Peninsula: Queries listed above each day indicate a higher feature score. Queries related to traffic, water, energy, logistics, and life reconstruction are highlighted in green, blue, yellow, orange, and pink, respectively.}\label{fig:result_jan_top15}
\end{minipage}}
\end{figure*}

Figure \ref{fig:result_jan_top15} summarizes the output results of the DNF for every day since December 31, 2023, the day before the disaster (until the end of January 2024). The top 15 queries with the highest scores are shown, but in reality, an average of about 100 information needs are output daily.

The results are colored for the five elements of Traffic, Water, Energy, Logistics, and Life Reconstruction. These colorings were not done automatically, but were done manually for the sake of clear explanation in this evaluation.
We will verify from two aspects that the results of this analysis successfully extracted the information needs of actual disaster areas.
The first is a comparison with actual news reports. It is shown that the information needs detected by DNF match with the news reports on the actual issues and concerns of disaster areas.
The second point is a comparison with other log data. While this study uses users' search query information and location information, we will show that the needs extracted by DNF are valid using other log data.

\subsection{Overall Result}
The increase or decrease in information needs in January 2024 for the five color-coded elements of Traffic, Water, Energy, Logistics, and Life Reconstruction is discussed over time. The details are shown below, but in general, the results indicate that DNF was able to accurately identify the information needs of the disaster areas.

\subsubsection{Traffic}
It has been observed since immediately after the disaster, and information needs related to traffic are strongly observed almost every day during the month of January. The Noto Peninsula has only one passable road, Noto Road, to drive into the peninsula from urban areas. The Noto Road was closed to traffic due to the earthquake, and was not re-opened until March 15. Therefore, while the Noto Road was closed, the Noto Peninsula remained an “isolated island on land” for a while, as relief supplies did not reach the Noto Peninsula sufficiently and people were unable to get out of these areas. This was of great concern to the residents of the affected areas, and there was a growing need for information on the state of restoration. Strong demand for information on gasoline and gas stations, which are indispensable for transportation, was also observed in the weeks immediately after the disaster.
In general, it can be read that information needs related to transportation remained high during the first month after the disaster.

\subsubsection{Water}

The water supply, which was damaged by the earthquake, has not been recovered in January, and therefore, information needs related to water are strongly observed on a daily basis. Furthermore, the information needs related to water were observed to change over time: for the first week after January 1, information needs related to “water outage” was the main need; after a week, information needs related to sanitation such as “toilets” and “bath” were observed. In the third week, information needs for “ Coin laundry” were increasingly observed. This indicates that the respondents wanted to do laundry, but were unable to do so, and as the days passed, the laundry accumulated, and in the third week, the need for laundry became stronger.
In any case, it can be read that they were always in a state of high information need for water as well.

\subsubsection{Energy}

As for electricity and gas, there had been a series of power outages immediately after the disaster, and it took about a week to restore them, so a strong need for information was observed immediately after the disaster. However, by the last week of January, many disaster sufferers were able to secure electricity at the minimum level, and the signals were almost no longer observed.
By examining the information needs of the DNF, it is possible to imagine, for example, how long the disaster areas were inconvenienced by the energy situation.

\subsubsection{Logistics}

As for logistics for the delivery of supplies, little was observed at the beginning, but the intensity of logistics became stronger around the third week. Since roads had not been restored in the disaster areas, the delivery of goods was difficult for some time. Sagawa Express was the first to gradually restore operations, and from mid-January, packages could be picked up only at its sales offices. Yamato Transport and Japan Post resumed deliveries in late January. Word of information needs have been observed in this order. This indicates that the information needs regarding deliveries in disaster areas were accurately picked up.

\subsubsection{Life Reconstruction}

Information needs related to rebuilding livelihoods have been observed almost daily since the second week of January.
In the first week, immediately after the disaster, people focused on adjusting to shelter life and securing water and food supplies, but from the second week onward, information needs related to disaster certification, insurance, bank accounts, and housing security were observed to be stronger.

\subsection{Validation1: Consistency of DNF Results with Media Reports}
\begin{figure}[t]
 \includegraphics[width=0.85\linewidth]{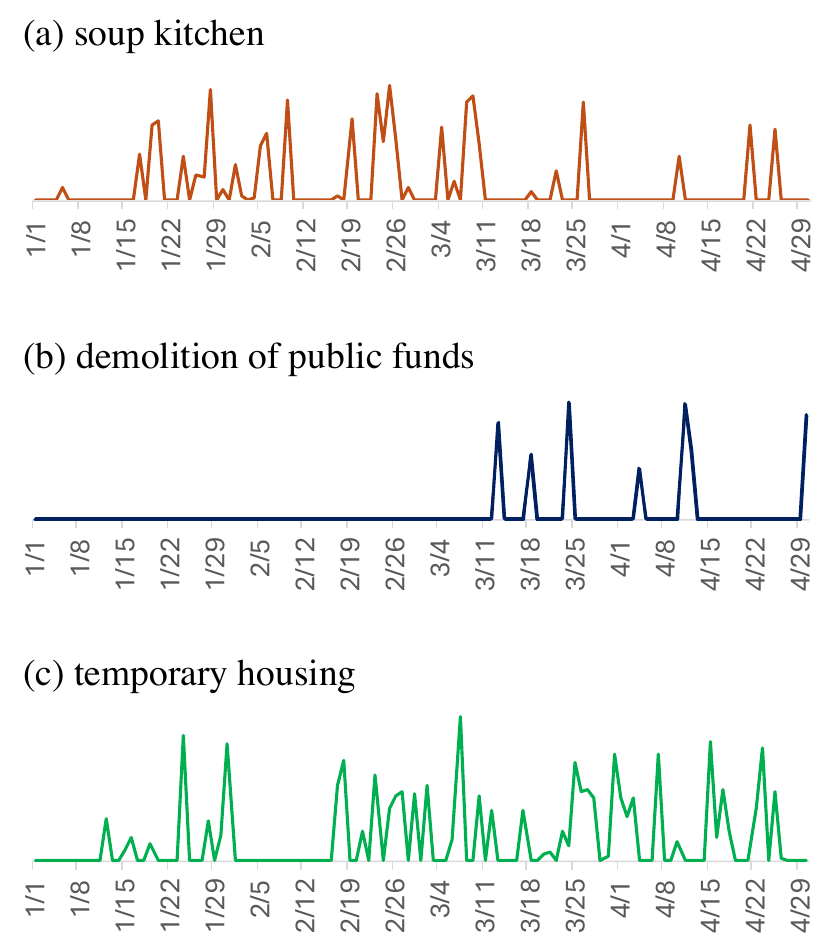}
 \caption{Trends in DNF scores over time for three information needs: soup kitchens, publicly funded demolition, and temporary housing}
 \label{fig:res1_add}
\end{figure}

The purpose of the analysis was to verify the validity of the information needs of the Noto earthquake disaster areas detected by DNF, and to check the consistency of the information needs with news reports. The period covered by the analysis was set until April 30, 2024. For the information needs that stood out among these needs, we searched for published media articles and checked their consistency.
As for the consistency between information needs and media that can be observed in the DNF scores since January, Figure \ref{fig:res1_add} separately shows the intensity by information need.

\subsubsection{Transition of Water Needs}
The transition from the need for water outage to the need for bath and toilet, and then to the need for coin laundry services, was also noted in the following article, which was reported on January 15 \footnote{\url{https://www3.nhk.or.jp/lnews/kanazawa/20240115/3020018179.html}}.

\begin{screen}
{\it A man in his 70s said, “I can't take a bath because the water is cut off and my head is itchy. I know this is a difficult situation, but I hope the water will be restored as soon as possible.
Another man in his 60s said, “I have trouble using the toilet because the water is not running, and I even went all the way to a laundromat in Kanazawa City to do my laundry. I am prepared for the situation to last a long time, but I hope you will do your best for the restoration.}
\end{screen}

\subsubsection{Logistics}
Logistic's information needs in mid to late January match up with the resumption of operations from Sagawa Express half a month later; an article published on January 12 also states the following \footnote{\url{https://www3.nhk.or.jp/lnews/kanazawa/20240112/3020018042.html}}:

\begin{screen}
{\it Sagawa Express resumed package pickup and delivery services in all areas of Nanao City and Nakanoto Town in Ishikawa Prefecture on January 12.
On the other hand, pickup and delivery of packages continues to be suspended in all areas of Wajima City, Suzu City, Anamizu Town, and Noto Town, respectively, and part of Shiga Town in Ishikawa Prefecture.}
\end{screen}

\subsubsection{Soup kitchen}
For disaster victims who cannot secure enough food, soup kitchens are a matter of life and death. Therefore, there is a high demand for information on takidashi throughout the period of the disaster. The fact that the Self-Defense Forces ended the soup-run on March 23 caused anxiety among the affected residents\footnote{\url{https://www3.nhk.or.jp/lnews/kanazawa/20240323/3020019635.html}}. As a result, since April, soup-runs have been conducted irregularly on a voluntary basis\footnote{\url{https://www3.nhk.or.jp/news/html/20240419/k10014427391000.html}}.

\begin{screen}
{\it The Self-Defense Forces began providing meals to evacuees in Suzu City on January 6, and three meals were served every day. However, the program will end on March 23.}
\end{screen}

\begin{screen}
{\it An event was held at an evacuation center in Nanao City, Ishikawa Prefecture, to interact with disaster victims.
At the meeting, a magician gave a show and volunteers served udon noodle soup.}
\end{screen}

\subsubsection{Publicly funded demolition}
Publicly funded demolition is a project to demolish or remove buildings damaged by the earthquake at public expense. Publicly funded demolition actually started in Noto Town on March 18, and Figure \ref{fig:res1_add}(b) shows that information needs regarding publicly funded demolition have been increasing in mid-March\footnote{\url{https://www3.nhk.or.jp/news/html/20240319/k10014396041000.html}}.
It was clear from the DNF scores that it was not appropriate from a safety standpoint to leave buildings about to collapse or piles of rubble where buildings had collapsed, and that publicly funded demolition was needed as soon as possible, but this was not easily accomplished due to road problems that prevented trucks from entering the demolition site.

\begin{screen}
{\it Noto Town has received applications from 167 people for demolition of damaged houses at public expense, and began demolition in the Ukawa and Shiromaru districts on March 18. The project will proceed toward completion in two years.}
\end{screen}

\subsubsection{temporary housing}
For temporary housing, demand for temporary housing began to be seen after 1/15 and will remain high after occupancy begins in early February. Information needs scores are also gradually increasing. As of March 1, applications for 7,800 units had been received and temporary housing was being constructed at a rapid pace. \footnote{\url{https://www3.nhk.or.jp/news/html/20240301/k10014376521000.html}}.

\begin{screen}
{\it According to Ishikawa Prefecture, as of the 22nd of last month, 7,800 units had applied to move into temporary housing.
Governor Hase said, “I would like to make every effort to ensure that disaster victims can move into temporary housing with peace of mind as soon as possible.}
\end{screen}

\subsection{Validation2: Consistency of DNF Results with Page View (PV) data}

\begin{figure}[t]
 \includegraphics[width=0.95\linewidth]{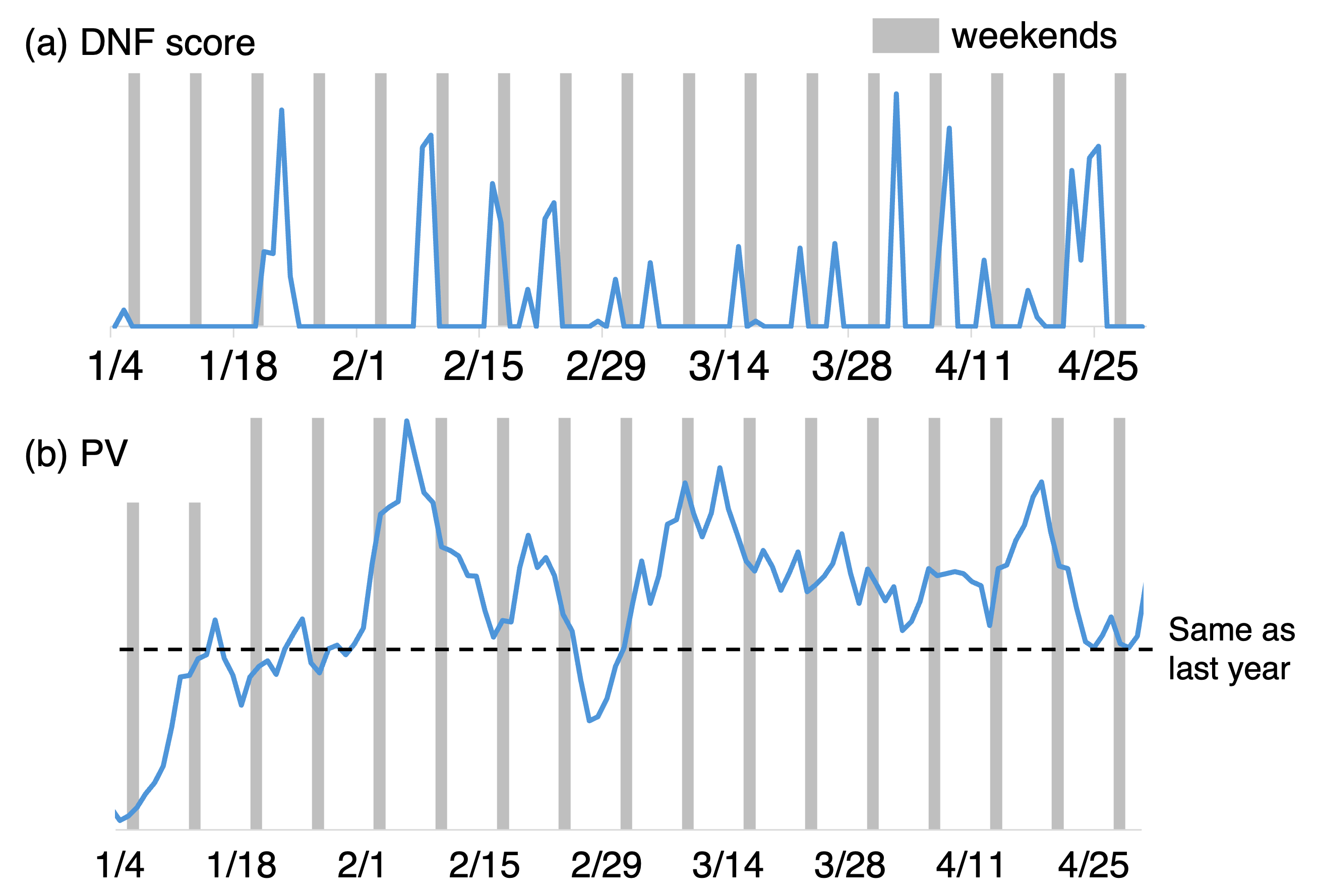}
 \caption{Trends in information needs for used cars: (a) shows the DNF score, and (b) shows the 7-day moving average of the PVs of used car websites compared to the previous year.}
 \label{fig:res2}
\end{figure}

Different methods from DNF are used to identify the needs of disaster areas to check the consistency of the results.
There are known methods for estimating the growing demand for a service by the number of PVs to the service's website, as in the study by Amit et al \cite{PVdemand}.
We will confirm the consistency between the existing method using PV and the DNF score proposed in this research.

\subsubsection{Survey Methodology}
This survey examines the subject of demand for information needs of used cars. When a disaster such as an earthquake occurs, there is an increase in demand for cars due to car breakdowns caused by the disaster and an increase in the need for cars for disaster recovery. In the case of the Noto earthquake, it was confirmed through on-site interviews that there was a large demand for cars at the site of the disaster, including free car sharing services, etc. Since new cars are not immediately available for purchase due to COVID-19 and the shortage of semiconductors, there is a large need for readily available used cars. Therefore, new cars were excluded from the study, and used cars were the focus of the survey.
LY Corporation also provides a used car-related information site called carview!\footnote{\url{https://carview.yahoo.co.jp/}}, and its PV volume can be monitored. The following discussion compares the amount of PV on the carview! site with the DNF score.

First, the DNF score was calculated in the same way as in Section 4.3 by extracting car-related words (Figure \ref{fig:res2} (a)).
Second, the PV score is a score indicating the intensity of the number of PVs observed on the used car pages of the carview! site compared to the same day in the previous year. The users who were used for the analysis of DNF score in the Noto Peninsula were selected as the target users. The PV score was calculated by aggregating the daily PVs of the target users from January 1, 2023, to April 30, 2024, and the results were compared with the previous year for the period from January 1, 2024, to April 30, 2024. 7-day moving average results are shown in Figure \ref{fig:res2} (b).

\subsubsection{Results}

According to Figure \ref{fig:res2}, we can confirm the consistency in the DNF score and PV score in the following four points, confirming the reliability of the DNF score.

First is the situation of {\bf no demand at the beginning of the disaster}. This assumes that people naturally cannot afford to consider purchasing a car immediately after the disaster.

Second, {\bf the increase in scores since late January} is seen in both graphs. In fact, reports of additional cars by donation have been seen during this period due to the shortage of car-sharing vehicles\footnote{\url{https://www3.nhk.or.jp/lnews/utsunomiya/20240209/1090016831.html}}. The demand for cars is increasing due to the increased demand for the transportation of debris and supplies.

Third, we can confirm the condition of {\bf intermittent score increases from February}. During this period, the shortage of cars was severe. This was a time when the car sharing association's free rental service was at full capacity due to reservations, and the lack of cars was often reported in the press\footnote{\url{https://www3.nhk.or.jp/lnews/kanazawa/20240224/3020019231.html}}.

Finally, {\bf both scores rise again at the end of April} can be read from both graphs: from the end of April to the first week of May, there is a long holiday weekend in Japan, and the demand for cars increased during this period\footnote{\url{https://www3.nhk.or.jp/news/html/20240425/k10014432951000.html}}.

\subsubsection{Resolution of DNF score}
In Figure \ref{fig:res2}, weekends are highlighted in gray.
Looking at (a), we can see a tendency for scores to increase on weekends or shortly before weekends, but this is not the case in (b). Since it has been confirmed from interviews with car sharing associations and other sources that demand for cars increases on weekends and holidays, we believe that DNF is a more timely estimator in this respect and more in line with local realities.

In general, the PV score includes as noise the demand blips caused by service campaigns and media coverage. It is not possible to accurately express demand simply by the number of PVs.
In addition, the DNF score shows signs of the actual increase on Saturdays and Sundays shortly before the weekend, and the ability to detect signs is an advantage of the DNF.

\section{Discussion}
\subsection{DNF Performance}
We evaluated the results of real-time DNF analysis of the Noto earthquake disaster for consistency with news reports and with the number of PVs of other services. The results showed that the DNF score was strongly consistent with other information, and was effective in synchronizing with the information needs of the disaster victims. The information needs of the disaster areas extracted by the DNF were actually reported in major media and web articles five days after the earthquake\footnote{\url{https://www3.nhk.or.jp/news/html/20240105/k10014309691000.html}}, enabling the rapid release of information to the public. The output of the DNF was reported several times afterwards in the major media and was well received\footnote{\url{https://www3.nhk.or.jp/news/html/20240122/k10014330111000.html}}, confirming the importance of analyzing information needs that change every moment of the day\footnote{\url{https://www3.nhk.or.jp/news/html/20240203/k10014346051000.html}}.

\begin{screen}
{\it LY Corporation's analysis of the characteristic words that were searched for more frequently in the Noto earthquake disaster areas revealed that the frequency of searches for words closely related to evacuation life, such as “water outage,” has changed over time.
The changes in characteristics obtained from the search query analysis appear to reflect the needs of disaster areas, which continue to require assistance according to their needs.}
\end{screen}

Many of the information needs that have been identified in the previous studies did not take into account the impact of media coverage and the information of people outside the disaster area. On the other hand, the proposed DNF succeeds in identifying local information needs with great precision through integrated analysis of location and search information, and machine learning to eliminate the influence of media reports and other factors.

\subsection{Motivation of estimating information needs}

Motivation for the DNF Framework will be discussed. In conclusion, the most important motivation is to clarify the information needs of disaster victims in order to provide appropriate support.
The situation in disaster areas changes from hour to hour, and the needs of those in the field change accordingly. The media also picks up information needs through interviews and reports on them, but they often emphasize one aspect of the situation. The information needs of disaster sufferers are diverse and change from moment to moment. In addition, there are information needs that can only be understood by those who are affected by the disaster, rather than being organized from an outsider's point of view. Therefore, it is very significant to provide a fair and comprehensive, data-driven approach to the diverse and rapidly changing information needs of disaster victims.

As we described in this paper, the proposed DNF is a completely data-driven approach, and it is possible to clarify the information needs that are truly desired in the field without preconceived notions. This can then lead to the provision of appropriate support and information from those around them.

\subsection{Applicability to other disasters}
Although this study focused its evaluation on the case of the Noto earthquake, the DNF Framework is naturally applicable to other disasters as well. The DNF Framework is effective not only for earthquakes, but also for typhoons, floods, heavy snowfalls, and other disasters that affect an entire area.
We would like to utilize the DNF Framework in other cases to help provide appropriate information in disaster areas.

Furthermore, the DNF framework can be applied to any region outside of Japan with location information and search history information. In this study, we assume that search history information contains signals of users' information needs, but other information is also applicable. We would like to demonstrate the DNF framework in other countries in the event of disasters.

\subsection{Limitation}
DNF uses machine learning to extract information needs through integrated analysis of location and search information. For simplicity, machine learning uses simple linear logistic regression. To increase the accuracy and resolution of information extraction, advanced machine learning methods such as neural networks and LLMs can be used.
Note, on the other hand, that the main proposal of this study is not a machine learning method but the DNF Framework and its demonstration, and the system is important. The accuracy of each part of the system and the efficiency of data processing can be expected to be improved by replacing it with other sophisticated methods.

\section{Conclusion}
We proposed a DNF framework for real-time observation of information needs in disaster areas, and demonstrated it in the Noto earthquake that occurred in Japan on January 1, 2024. The DNF was actually put to work in the Noto earthquake, providing information needs on a continuous basis, and was highly evaluated as a new approach to support disaster areas. Future work will be conducted to demonstrate the effectiveness of DNF in other disasters as well.

%\newpage
\bibliographystyle{ACM-Reference-Format}
\bibliography{references}

\end{document}